\documentclass[a4paper,10pt,onecolumn]{article}
\usepackage{amsmath}%
\usepackage{amsfonts}%
\usepackage{amssymb}%
\usepackage{graphicx}
\usepackage{array}
\usepackage{color}
\usepackage{float}
\usepackage[font={small,it}]{caption}
\usepackage[letterpaper, margin=1in]{geometry}
\usepackage[T1]{fontenc}
\usepackage[utf8]{inputenc}
\usepackage{authblk}
\usepackage{subcaption}

\begin{document}

\title{Localized fluidization in granular materials: Theoretical and numerical study}
\author[1]{E. P. Montell{\`a}}
\author[1]{M. Toraldo}
\author[1]{B. Chareyre}
\author[1]{L. Sibille}
\affil[1]{University Grenoble Alpes (UGA), 3SR, F-38000 Grenoble, France}

\date{9 August, 2016}
\maketitle

\section*{Abstract}
We present analytical and numerical results on localized fluidization within a granular layer subjected to a local injection of fluid. As the injection rate increases the three different regimes previously reported in the literature are recovered: homogeneous expansion of the bed, fluidized cavity in which fluidization starts developing above the injection area, and finally the chimney of fluidized grains when the fluidization zone reaches the free surface. The analytical approach is at the continuum scale, based on Darcy's law and Therzaghi's effective stress principle. It provides a good description of the phenomenon as long as the porosity of the granular assembly remains relatively homogeneous, i.e. for small injection rates.
The numerical approach is at the particle scale based on the coupled DEM-PFV method. It tackles the more heterogeneous situations which occur at larger injection rates. The results from both methods are in qualitative agreement with data published independently. A more quantitative agreement is achieved by the numerical model. A direct link is evidenced between the occurrence of the different regimes of fluidization and the injection aperture. While narrow apertures let the three different regimes be distinguished clearly, larger apertures tend to produce a single homogeneous fluidization regime. 
In the former case, it is found that the transition between the cavity regime and the chimney regime for an increasing injection rate coincides with a peak in the evolution of inlet pressure.
Finally, the occurrence of the different regimes is defined in terms of the normalized flux and aperture.

\section{Introduction}

In a broad sense, fluidization refers to the fluid-induced mobility of solid grains in a granular material subjected to upward seepage flow \cite{payne2008remediation}. Fluidization is employed in a wide variety of industrial processes such as heat transfer, petroleum refining, coal conversion and water treatment \cite{peng1997hydrodynamic,weisman1994design}. It may also occur as a result of seepage flow in a soil, in which case it can be the cause of internal soil erosion that can lead to serious failures of hydraulic works (dykes, levees, dams, etc) \cite{bonelli2013erosion,foster2000statistics,fry1997erosion}. A particular case is when there is very localized influx of fluid, leading to a spatial heterogeneity of the phenomenon, this situation is generally termed \textit{localized fluidization} in the literature. Such a configuration appears in tapered fluidized bed reactors found in many industrial process (drying, coating crystallization, mixing, etc) \cite{schaafsma2006investigation,
sutar2012mixing}, spouted beds, or in some natural geological formations \cite{sutkar2013spout}. Eventually, fluidized zones induced by underground pipe leakage are also a major concern 
as ground surface may collapse due to the leak, causing important accidents \cite{soderlund2007evaluating}. In addition, channelling can be observed in some applications of fluid bed reactors. Channelling is a condition wherein the fluid passes through the bed along localized paths \cite{briens1997characterization}. This phenomenon should be avoided due to its adverse effects. Hydro-mechanical instabilities have been observed experimentally and simulated numerically in the case of a saturated granular medium when a localized flux is injected through a small orifice \cite{gallo2004steady,kohl2014magnetic,philippe2013localized,zoueshtiagh2007effect}.

Despite the large number of works dedicated to fluidized beds \cite{anderson1967fluid, gidaspow1991hydrodynamics,mickley1955mechanism}, only a few have focused on the initial and developing phases of a localized fluidized zone inside a granular medium \cite{cui2014coupled, ngomainteraction,philippe2013localized,zoueshtiagh2007effect}. The present study is devoted to this specific aspect.

\begin{figure}[H]
    \centering
        \includegraphics[width=10cm]{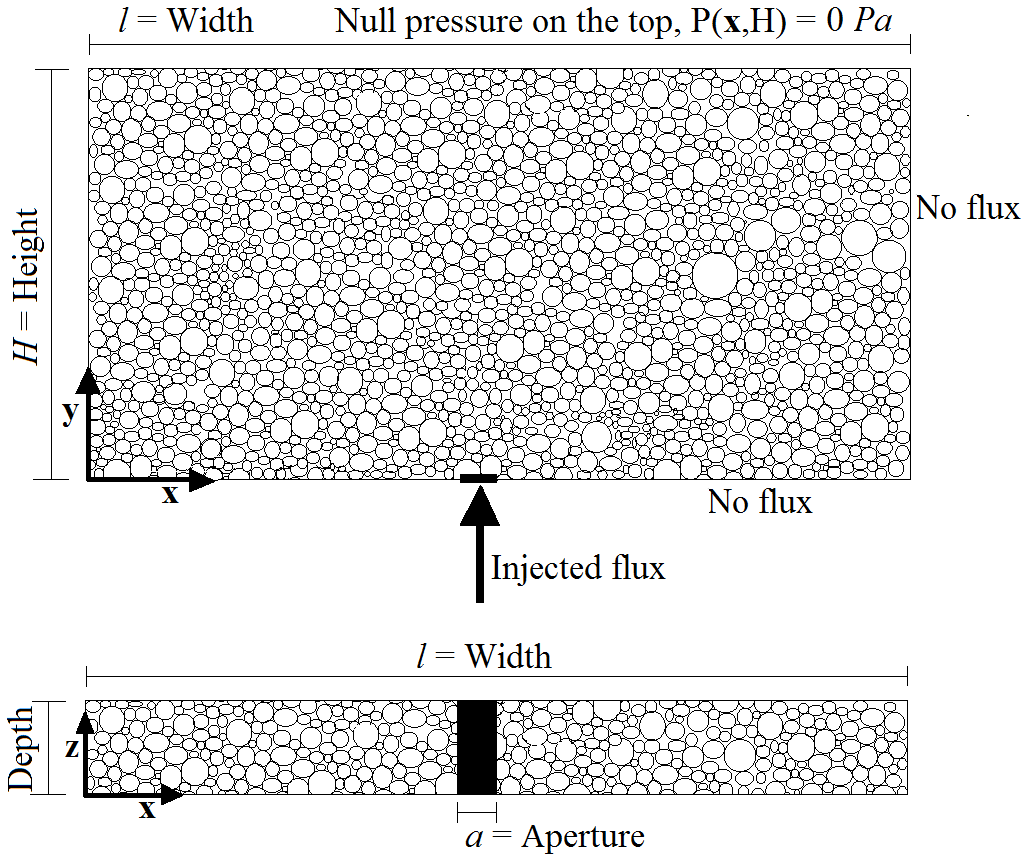}   
    \tiny
    \caption{Sample geometry and boundary conditions.}
    \label{new_figures:studycase}
\end{figure}

A typical configuration for laboratory experiments on localized fluidization is shown in figure \ref{new_figures:studycase}. The previous works on such configurations evidenced three successive regimes during a gradual increase of the injection rate \cite{ngomainteraction,philippe2013localized,zoueshtiagh2007effect}. At very low rates, the bed is stable. Larger rates cause bed expansion even before any  fluidization zone can be observed (expansion regime). For a yet larger rate, the hydrodynamic forces exerted on some particles are sufficient to counterbalance their weight, triggering movements above the injection point in the so called "fluidized" zone (cavity regime). Eventually, the height of the fluidized zone increases with the injection rate, until it reaches the top of the granular layer (chimney regime). Experimental results on the progressive development of a fluidized zone in a saturated bed of grains under the effect of a localized upward flow \cite{philippe2013localized} enabled to gain insight into the fluidized regimes. However, information at the grain scale and details of the pressure field were not accessible by this way. In order to overcome these limitations and analyze more deeply the local mechanisms responsible for fluidization and grain destabilization a numerical model, embedding fluid-solid coupling on the micro-scale, is used in this work.

Overall, the process is highly non-linear, which makes the analytical study impossible without crude simplifications. Nevertheless, it will be seen that a continuum scale model can give some insight into some governing mechanisms. In this work, we provide closed form solutions for the field of initial pore pressure in the problem of figure \ref{new_figures:studycase}. It enables the determination of the effective stress at every point of the problem. Then the effective stress field will be used to determine the extents of the fluidized zone.

A more realistic simulation of the process is possible using a numerical model of grain-fluid systems. To this aim, we use a coupling between the discrete element method and a pore-scale finite volume method (DEM-PFV) \cite{catalano2014pore,chareyre2012pore} to simulate the complete process from the expansion regime to the chimney regime. 

This work is devoted to two objectives. Firstly, analytical and numerical simulations performed in this study enable a better understanding of the local mechanisms responsible for fluidization and which variables govern the phenomenon describing the initial and developing phases of fluidization from punctual to a source of infinite size and completing the experimental previous work \cite{philippe2013localized}. Secondly, by tracking the effective stress within the medium it is shown that the effective stress constitutes a relevant parameter assessing the occurrence of fluidization. Thus, the effective stress field, as well as the porosity field, gives a good approach to describe the internal configuration and define the fluidized zones inside the granular assembly.

The paper is organized as follows: First, we define the physical model and the relevant dimensionless variables. Secondly, the theoretical and numerical models are introduced in section 2. The results are presented in section 3 and compared to data available in the literature. Finally, numerical results are used to highlight the role of flow rate, particle sizes, aperture of the injection zone, and viscosity of the pore fluid. The domains corresponding to the different fluidization regimes are defined in terms of dimensionless injection rate and dimensionless aperture. 

\section{Methodology}
\subsection{Problem statement and experimental set-up}\label{data_analytical}

As a model system we consider a layer of mono-disperse spheres immersed in a viscous fluid inside a rigid box. The system is subjected to gravity and the density of the solid particles is larger than that of the fluid (i.e. the particles sink). The granular layer is initially at static equilibrium then subjected to a local injection of fluid through the bottom face of the box (figure \ref{new_figures:studycase}). At the free surface of the granular layer the total stress and the fluid pressure are both null. The lateral and bottom faces are fixed (i.e. zero-displacement condition) and impermeable - the injection orifice excepted. The injection occurs through a rectangular area that covers the whole depth of the box. In this injection area the boundary condition for the solid particles is the same as for the rest of the bottom face: no displacement, as if a rigid grid was stopping the particles while letting the fluid pass through.

Hereafter, the theoretical analysis of this problem is two-dimensional (2D), thus exploiting the invariance along depth when the distance between the front and back plates is much smaller than the other extents (as found in the published data). On the other hand, the numerical model considers three-dimensional (3D) sphere assemblies. Granular systems are indeed 3D at the micro-scale even when the average displacement field is 2D. It was thus considered a requirement to simulate three-dimensional grains to approach realistic responses. Periodic boundary conditions are assumed along the horizontal directions, consistently with the two-dimensionality at the macroscale. They are preferred over rigid faces in order to not introduce heterogeneities of the microstructure near the boundaries.

The setup described here is inspired by the physical experiments of \cite{philippe2013localized}. Spherical beads were poured into a box before the box was slowly filled with oil via the bottom injection hole. The experiments were carried out by imposing different injection rates. Flow through the porous medium remained in the Stokesian regime as Reynolds number was kept low during the experiments. Stokesian regime is considered in the theoretical and numerical models as well. Visualization of the granular structure was made possible by the combined use of two optical techniques: refractive index-matching between the liquid and the beads and planar laser-induced fluorescence. Fluidization - in fact particles mobility - was then evaluated by image processing. The injection was done through a circular aperture of diameter $D$ = 14 mm at the center of the bottom face. We will not consider a circular aperture in our model system in general since it breaks the invariance along depth, thus introducing additional and superfluous complexity. However, one numerical simulation with a circular aperture has been carried out and will be reported for direct comparisons with the experimental result.

The physical variables of the problem are summarized in table \ref{tab:Solid_fluid_properties}.

\begin{table}[H]
    \centering
    \small{\begin{tabular}{|p{40mm}|p{16mm}|p{16mm}|}
            \hline
            \textbf{Variable} & \textbf{Dimension} & \textbf{SI units}\\
            \hline
              Discharge per unit depth ($q$) & $L^{2}\cdot T^{-1}$ & $[m^2/s]$\\
            \hline
              Pressure ($P$) & $FL^{-2}$ & $[Pa]$\\
            \hline
              Height ($H$)  & $L$&  $[m]$\\
            \hline
              Length ($l$)  & $L$&  $[m]$\\
            \hline             
              Grain diameter ($D$) & $L$& $[m]$\\
            \hline
              Dynamic viscosity ($\mu$)   & $FL^{-2}T$& $[Pa\cdot s]$   \\
            \hline
              Aperture($a$)   & $L$& $[m]$   \\
            \hline
              Solid eight density ($\gamma_s$)   &  $FL^{-3}$ & $[N\cdot m^{-3}] $\\
            \hline
              Fluid weight density ($\gamma_w$)   &  $FL^{-3}$ & $[N\cdot m^{-3}] $\\
            \hline       
              Porosity ($n$)  & $-$ & $[-]$ \\
            \hline
    \end{tabular}
    }
    \caption{Physical and geometrical variables of the problem. The dimensions are defined in the [FLT] (force, length, time) system.}
    \label{tab:Solid_fluid_properties}
\end{table}

The porosity of the porous medium appearing in table \ref{tab:Solid_fluid_properties} is defined by the ratio $n=\dfrac{V_v}{V_t}$, where $V_v$ is the volume of void-space and $V_t$ the total volume of material (note that $n=1-\phi$ if $\phi$ denotes the solid fraction).

The physical properties of the materials used in the experiments are specified in table \ref{tab:sampleCharacteritics}:

\begin{table}[H]
    \centering
    \small{\begin{tabular}{|p{42mm}|p{18mm}|}
            \hline
            \textbf{Characteristic} & \textbf{Experiment}  \\
            \hline
               Width ($L$)    & 0.20 $m$ \\
            \hline
              Initial height ($H_o$)  & 0.12 $m$ \\
            \hline
              Depth ($s$)   & 0.08 $m$\\
            \hline             
              Mean radius ($r_m$)   & 0.00250 $m$\\
            \hline
              Density of the solid phase ($\rho_s$)   & 2230 $kg\cdot m^{-3} $ \\
            \hline
              Density of the fluid phase ($\rho_f$)   & 850 $kg\cdot m^{-3} $ \\
            \hline             
              Dynamic viscosity ($\mu_s$)   & 0.0183 $Pa\cdot s $ \\
            \hline
    \end{tabular}
    }
    \caption{Solid and fluid properties of the experiments.}
    \label{tab:sampleCharacteritics}
\end{table}

\subsection{Dimensionless variables}\label{Dimensionless_analysis}

Based on the above variables we introduce the so-called submerged (or apparent) density of the solid phase $\gamma'=(1-n)(\gamma_s-\gamma_w)$, and the reference vertical effective stress $\sigma'_0$ corresponding to the intergranular stress at the bottom of the layer (see next section), i.e.
\begin{equation} \label{eq:refSigma} \sigma'_0=(1-n)(\gamma_s-\gamma_w)H= \gamma'H. \end{equation}
Normalization by this reference pressure leads to the following dimensionless group, where the normalized form of each variable is denoted by the ``$*$''. The relevance of this set of dimensionless variables will be demonstrated in section 4. Note that the dimensionless fluid pressure is  a normalized \textit{excess} pore pressure, i.e. the difference between absolute pressure and hydrostatic pressure.

\begin{itemize}
 \item Normalized fluid pressure 
 \begin{equation} \label{eq:Ec6} \qquad p^*=\dfrac{P+\gamma_w(y-H)}{\sigma'_0}   \end{equation}
 \item Normalized flux
 \begin{equation} \label{eq:Ec4} \qquad q^*= \dfrac{q\mu}{D^{2}\sigma'_0}   \end{equation}
 \item Normalized coordinates
 \begin{equation} \label{eq:Ec7} \qquad x^*=\dfrac{x}{l}  \end{equation}
 and 
 \begin{equation} \label{eq:Ec8} \qquad y^*=\dfrac{y}{H}  \end{equation}

 \item Normalized aperture
 \begin{equation} \label{eq:Ec8.5} \qquad a^*=\dfrac{a}{l} \end{equation}

\end{itemize}

As it will be shown later, the response of the system for a given injection rate strongly depends on the macro-scale hydraulic conductivity $K$ of the granular material (ratio between seepage velocity and pressure gradient). $K$ is proportional to the squared particle size as in $K=\kappa_0 \dfrac{D^{2}}{\mu}$, where $\kappa_0$ is dimensionless and depends on porosity only (see e.g. the Kozeny-Carman form of this relationship). Through $K$ there is an effect of particle size in the continuum scale modeling. An alternative definition of the normalized flux is thus, instead of Eq.\ref{eq:Ec4}, 
\begin{equation}\label{eq:Ec5}  q^*_k= \dfrac{q}{K \sigma'_0}. \phantom{\hspace{3.4cm}} \end{equation}
Both $q^*$ and $q^*_k$ will be used in the analysis. The definition of $q^*$ is simpler and, since $K$ has not been measured in the experiments of \cite{philippe2013localized}, it is the only form we can use for comparing data and simulations. $q^*_k$ has the advantage of reflecting the change of porosity at any step of fluidization and will be used for the interpretation of some numerical results.

\subsection{Theoretical model}\label{analytical_solution}

We suggest that fluidization can be seen as a special case of the so called \textit{liquefaction}. The latest refers to situations in which the total stress tensor $\sigma$ in a saturated material and the pressure of the pore fluid $P$ are such that the \textit{effective stress} tensor
\begin{equation}
\label{effective}
 \sigma'=\sigma+P\boldsymbol I
\end{equation}
vanishes or has at least one vanishing eigen value. In this context, fluidization is simply the liquefaction produced by a particular combination of gravitational acceleration and upward seepage flow (while liquefaction in general can occur in non-gravitational systems and regardless of seepage flow).

Momentum balance for the solid-fluid mixture in the Stokesian regime lets one deduce the component $\sigma_{yy}$ of the total stress in the problem 
\begin{equation}
 \sigma_{yy}=\gamma_{sat}(y-H),
\end{equation}
where $\gamma_{sat}$ is the average weight density of the saturated material, defined by $\gamma_{sat}=n\gamma_w+(1-n)\gamma_s=\gamma_w+\gamma'$.

As long as the free surface is approximately flat and the porosity (hence $\gamma_{sat}$) is approximately uniform, $\sigma_{yy}$ is constant. Consequently,   
the changes in the effective stress in Eq. \ref{effective} can only result from a change of the excess pore pressure, itself controlled by the injection rate.
The key part of the theoretical modeling is thus to determine the spatial distribution of pore pressure within the specimen to identify the zones in which it reaches (or exceeds) the total stress.

The granular material will then be considered fluidized at a particular location $(x,y)$ if
\begin{equation}
 \gamma_{sat} \cdot (y-H)+P(x,y) \geq 0,
\end{equation}
or in an equivalent dimensionless form, introducing the normalized effective stress $\sigma'^*$:

\begin{equation}
\label{fluidizationThreshold2}
 \sigma'^* := \dfrac{y-H}{H}+p^*(x,y) \geq 0,
\end{equation}

In order to find closed form solutions for $P$, the following assumptions have been considered:
\begin{itemize}
  \item  Darcy's law applies at the bulk scale, i.e. the seepage flow is driven by the gradient of excess pore pressure with a velocity $\boldsymbol v=-K\nabla(P+\gamma_wy)$.
  \item  the porous medium is homogeneous and the deformation is null or negligible, hence conductivity $K$ is uniform in space and time.
\end{itemize}

Under these assumptions, the final expression of the pressure at any point of the specimen can be obtained as a sum of the pressures induced at this point by an infinite set of punctual sources/sinks which symmetries replicate the actual boundary conditions (see appendix for further details):

 \thickmuskip=0mu
 \makeatletter
 \def\@eqnnum{{\normalsize \normalcolor (\theequation)}}
  \makeatother
 { \small \begin{equation}\label{eq:Ec3}  P+\gamma_wy=\dfrac{q}{2 \pi K} \sum\limits_{j=-\infty}^\infty \sum\limits_{i=-\infty}^\infty -1^{|j|} \left[ln(\sqrt{(x-i\ l)^{2}+(y-jc)^{2}}) \right]
 \end{equation} }
where $c=2H$ as sources and sinks are spaced by a distance equal to twice the sample height.
If a finite injection area is considered rather than an injection point, the above expression needs to be integrated on the aperture width:

 \thickmuskip=0mu
 \makeatletter
 \def\@eqnnum{{\normalsize \normalcolor (\theequation)}}
  \makeatother
 { \small\begin{equation}\label{eq:Ec3.2} P+\gamma_wy=\dfrac{q}{2 \pi K} \int_{-a/2}^{a/2} \sum\limits_{j=-\infty}^\infty \sum\limits_{i=-\infty}^\infty -1^{|j|} \left[ln(\sqrt{(x-i\ l-s)^{2}+(y-jc)^{2}}) \right] ds
 \end{equation} }
 
Note that the pressure defined by Eq. \ref{eq:Ec3} is singular at the injection point, where the pore pressure is infinite. In the limit $a^*=0$ there is therefore a finite-sized fluidized zone for any value of the injection rate. Conversely, Eq. \ref{eq:Ec3.2} takes finite values in the injection area, thus defining a clear threshold in terms of injection rate below which the effective stress is strictly compressive everywhere.

According to Eq. \ref{eq:Ec3.2}, $K$ must be known in order to find the pore pressure. In the DEM-PFV numerical model, $K$ is a result, and it depends on the particle size and the porosity of the assembly. Consequently, the analytical and the numerical methods can be compared directly by plugging $K$ from the numerical model into Eq. \ref{eq:Ec3.2}.

It is important to recall, however, that Eq. \ref{eq:Ec3.2} is only valid for a homogeneous medium, which is not necessarily the case in experiments or numerical simulations. As soon as localized fluidization occurs, significant differences are expected between this equation and the actual or simulated fluid pressure.


\subsection{Numerical model}\label{Numerical_method}

The behavior of a granular bed subjected to localized upward flux can be investigated using different coupled DEM-fluid models. Some previous works were based on couplings between the DEM and the Lattice Boltzmann \cite{chen1998lattice} (LBM) method in two dimensions \cite{cui2014coupled, ngomainteraction,cui20122d,cui2013numerical}. 

2D granular systems are peculiar from a mechanical point of view and their hydraulic properties are unclear - strictly speaking they are impermeable since they don't offer any free path to the fluid.
A quantitative approach of the problem thus requires 3D models. To this aim, the so-called DEM-PFV coupling was used for the present study \cite{catalano2011pore}. DEM-PFV refers to a micro-hydromechanical model combining the DEM and a pore scale finite volume formulation of the viscous flow of an incompressible pore fluid \cite{chareyre2012pore,catalano2014pore}. It enabled 3D simulations at a reduced cost compared to 3D DEM-LBM simulations.

The solid particles in the model are spherical and slightly poly-disperse (uniform distribution deviating by 2\% from the mean diameter). The interaction between them are elastic-plastic, with normal and tangential stiffness $k_{n}$ and $k_{s}$, and Coulomb friction angle $\phi$. Newton's second law of motion is integrated explicitly through iterative time-stepping (implementation details can be found in \cite{yade:doc2}). The fluid flow model is based on a pore scale discretization of Stokes equations, where the pores are defined by the tetrahedra of a \textit{regular triangulation} \cite{chareyre2012pore}. At each time step, the geometry and rate of deformation of each pore is updated on the basis of particles motions. In turn, the fluxes are determined and the fluid forces on the particles are obtained. They are integrated in the law of motion for each particle. This work is carried out using the PFV implementation provided by the open source code YADE-DEM\cite{yade:doc2}. 

The initial granular layer is obtained by simulating the gravitational deposition of a cloud of particles in a periodic box. The deposition stage stops when the particles reach static equilibrium. The layer is then subjected to an influx of fluid at the bottom, as shown in figure \ref{new_figures:studycase}. The simulated granular layer is made up of 5000 spheres, the mean diameter of the grain is $D = 0.0166m$ and the height of the sample is $H \approx 19D$.

The effective stress in simulated granular systems can be computed directly based on the contact network. Following \cite{catalano2014pore}, the average effective stress tensor associated to one particle of a saturated material is defined by
\begin{equation}
\label{eq:effectiveS}
 \sigma'=\dfrac{1}{V_p}\sum_{k=1}^{N_c}\boldsymbol f_k \otimes \boldsymbol x_k,
\end{equation}
where $N_c$ is the number of contact with other particles, $\boldsymbol f_k$ is a contact force, $\boldsymbol x_k$ the position vector of the contact point, and $V_p$ the volume of the Voronoi cell enclosing the particle. This definition results in rather scattered values when plotted per particle but meaningful results can be obtained when they are averaged locally (see section \ref{numerical_results}).


\section{Results and discussion}
\subsection{Analytical solution}\label{Analytical_solution}

\begin{figure}[H]
    \centering
        \includegraphics[width=8cm]{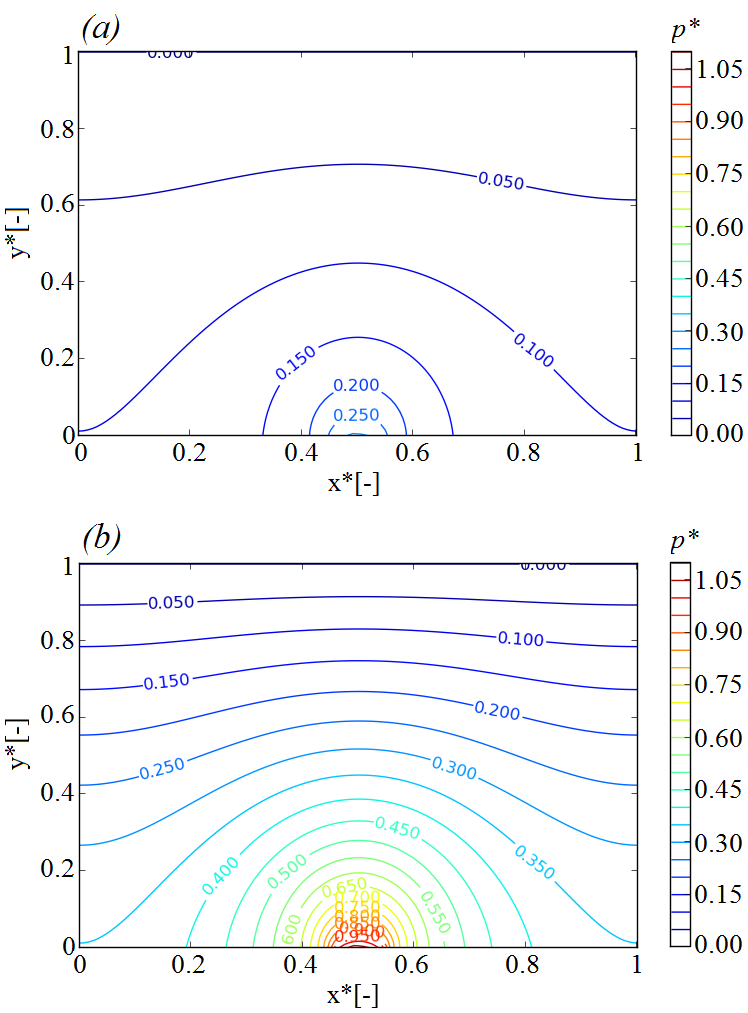}   
    \caption{Evolution of the dimensionless pressure field $p^*$ for an injection aperture $a^*$ = 0.1 and two injection rates $q^*$ = 0.00019  (a) and $q^*$ = 0.00067 (b).}
    \label{new_figures:pressure_field}
\end{figure}

\begin{figure*}[ht]
    \centering
        \includegraphics[width=17cm]{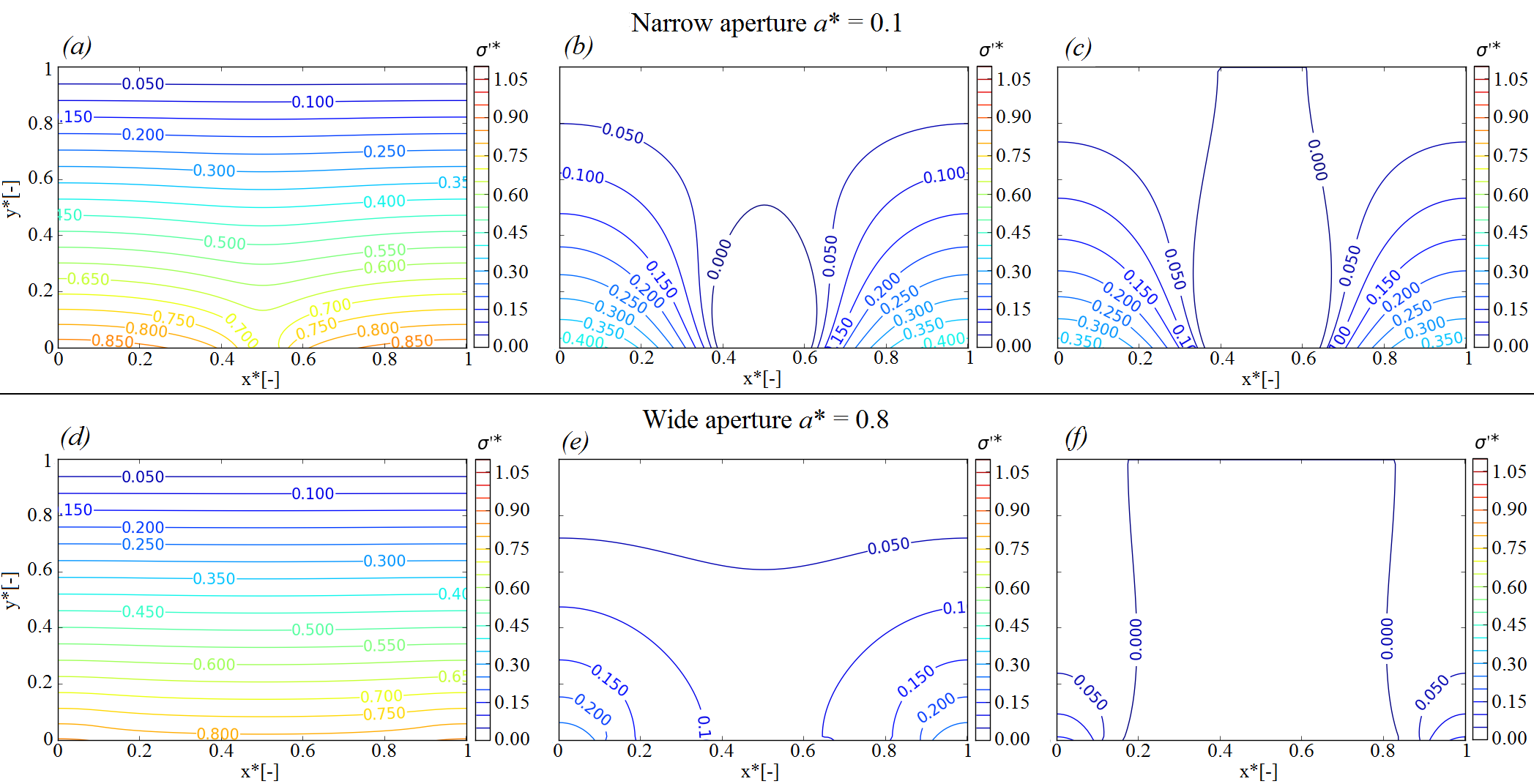}   
    \caption{Evolution of the dimensionless effective stress field when the injected flux increases. Narrow aperture ($a^*=0.1$) on the first line and wide aperture ($a^*$ = 0.8) on the second line. The normalized fluxes are $q^*$ = 0.00019 (a,d), $q^*$ = 0.00105 (b,e), $q^*$ = 0.00114 (c) and $q^*$ = 0.00126 (f).}
    \label{new_figures:effective_field}
\end{figure*}

Figure \ref{new_figures:pressure_field}  shows the evolution of the pressure field after Eq. \ref{eq:Ec3.2} for increasing flux values and an aperture of 10$\%$ of the width of the specimen ($a^*=0.1$ (Eq.\ref{eq:Ec8.5})). Near the injection area, the pressure contours tend to concentric half-circular shapes due to the quasi-radial flow distribution. On the other hand, the isolines are horizontal near the side walls, consistently with the no-flux condition.
Following section \ref{analytical_solution}, the fluidized zone can be identified with respect to the sign of the effective stress. In order to show the evolution of the fluidized zone and to identify the flux values triggering different steps of the fluidization phenomenon, the normalized effective stress is plotted in figure \ref{new_figures:effective_field} for apertures $a^*$ = 0.1 (plots $(a)$, $(b)$ and $(c)$) and $a^*$ = 0.8 (plots $(d)$, $(e)$ and $(f)$). Therein the extent of the fluidized zone can be defined through the shape of the null-pressure isoline.

From the first series of plots (narrow aperture) we can distinguish the regimes described in \cite{philippe2013localized}. Low flux values correspond to the expansion regime in which no fluidization zone is detected (plot $(a)$ in figure \ref{new_figures:effective_field}).

As the flux increases, pore pressure keeps building up until it balances the total stress. At this point a fluidized zone starts developing above the injection area, corresponding to the cavity regime (plot $(b)$ in figure \ref{new_figures:effective_field}). Eventually, the fluidized zone reaches the top of the specimen, leading to a chimney of fluidized grains (plot $(c)$ in figure \ref{new_figures:effective_field}).

The solution with a wider injection area (bottom series, plots $(d)$, $(e)$ and $(f)$ in figure \ref{new_figures:effective_field}) shows that a slightly larger flow rate is required to initiate the fluidization above a wide aperture.

Distinguishing the cavity regime from the chimney regime in this situation is made uneasy by the fact that large aperture tends to fluidize the granular layer simultaneously at every point in space, thus merging the cavity and chimney regimes into one single regime as $a^*\rightarrow 1$.


\subsection{Numerical simulations}\label{numerical_results}

\begin{figure}[H]
    \centering
        \includegraphics[width=16cm]{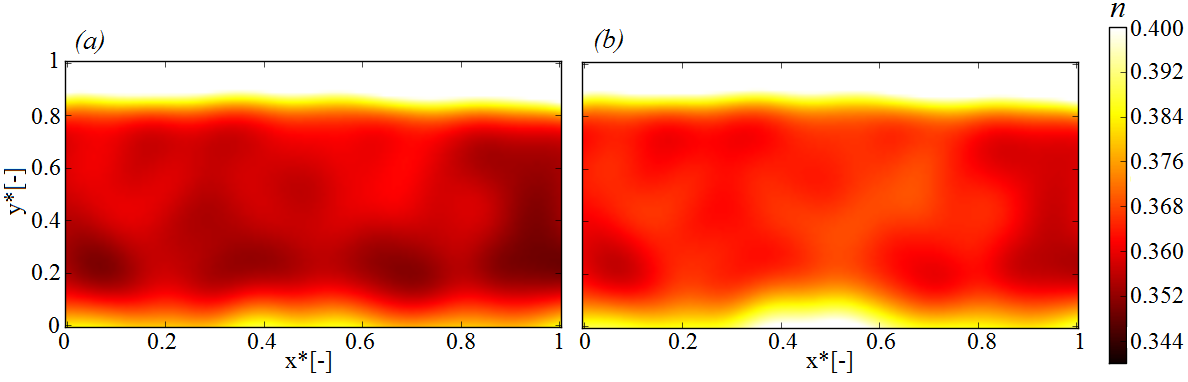}  
\caption{Evolution of the porosity for narrow aperture ($a^*$ = 0.1) in a very detailed color scale. (a) Static regime, $q^* $= 0. (b) Expansion regime, $q^*$ = 0.00095.} \label{new_figures:porosity_expansion}
\end{figure}

\begin{figure}[H]
    \centering
        \includegraphics[width=16cm]{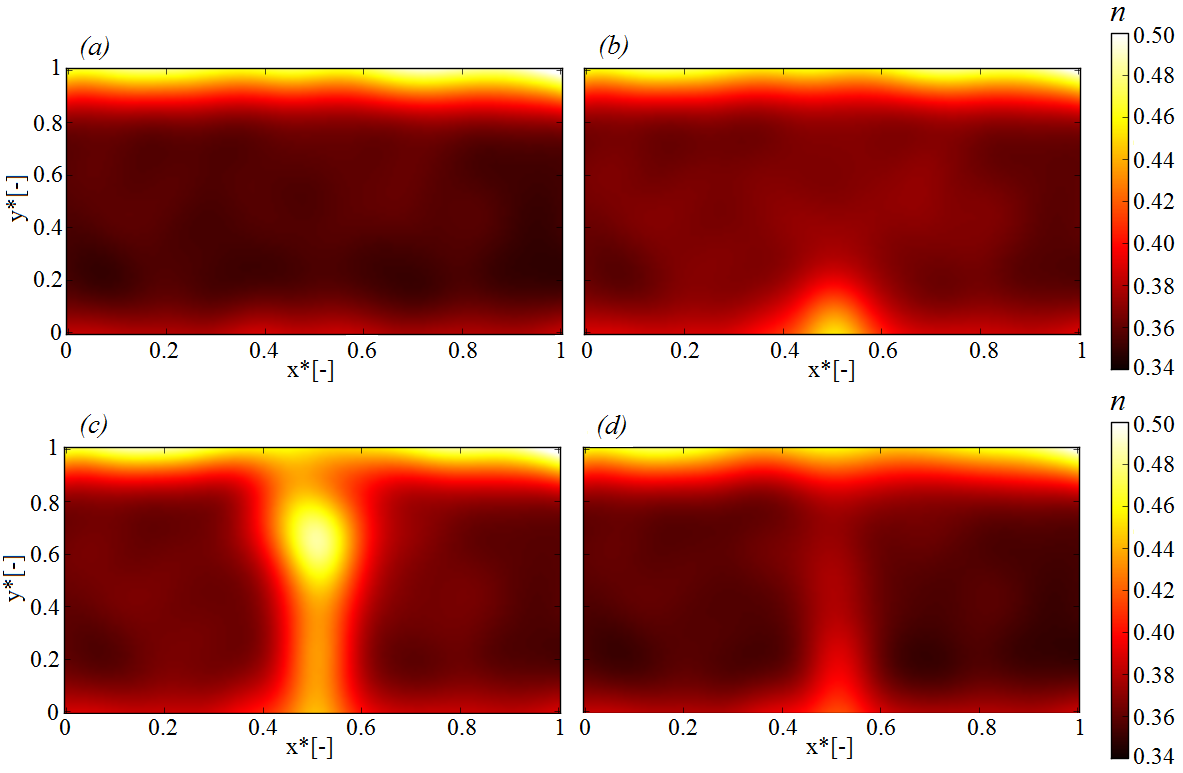}   
\caption{Evolution of the porosity for narrow aperture ($a^*$ = 0.1). (a) Static regime, $q^*$= 0. (b) Cavity regime, $q^*$ = 0.00116. (c) Chimney regime, $q^*$ = 0.00145. (d) Irreversible changes in porosity after reducing the flux back to $q^*$ = 0.} \label{new_figures:porosity_cavity}
\end{figure}

Figure \ref{new_figures:porosity_expansion} and  \ref{new_figures:porosity_cavity} show the evolution of porosity within the simulated layer. Porosity is defined for each particle as the ratio of the volume of the void to the total volume of the Voronoi cell enclosing the particle \cite{catalano2014pore}. The fields of figures \ref{new_figures:porosity_expansion} and \ref{new_figures:porosity_cavity} are in fact a moving average of the per-particle porosity for reducing the noise due to the scattered local values. The large porosity values at the top of the layer are artifacts and should be disregarded: the particles of the free surface have large porosity values by definition as their Voronoi cells enclose some void space above the free surface. This artifact gives a fringe of high porosity through the moving average procedure. We notice that, with the exception of this top boundary layer, the deposition process results in a rather uniform porosity throughout the sample. To some extent a high porosity layer is also noticeable near the bottom plate, in this case not an artifact but a perturbation of the microstructure by the rigid wall. There is no perturbation near the vertical boundaries thanks to the periodicity.

The porosity changes inside the layer are shown in figure \ref{new_figures:porosity_expansion} and \ref{new_figures:porosity_cavity} for $a^*=0.1$. A narrow color scale is used in the former to emphasize the expansion regime, which appears as a rather homogeneous process. During the expansion regime the porosity increases over all the points within the sample (from $n=0.360$ to $n=0.368$ in average, excluding the free surface) and no significant heterogeneities are detected.
The height of the granular layer increases almost uniformly during this expansion, reaching values over $y^*=1$ (see figure \ref{new_figures:experimental_DEM_curves}).

Figure \ref{new_figures:porosity_cavity} highlights the heterogeneous changes of porosity as localized fluidization starts developing: a cavity appears for $q^*=1.16\times10^{-3}$, followed by a chimney for $q^* \geq 1.45\times10^{-3}$.
In both cases the particles located in the regions of low porosity are moving and have only transient contacts with each other, while the particles of the dense regions are static and contribute to a permanent contact network. 


Plot $(c)$ in figure \ref{new_figures:porosity_cavity} ($q^* = 1.45\times10^{-3}$) shows inside the chimney a region of even higher porosity ($n\approx 0.5$ near $y^*=0.7$). This pattern is typical and was found in most simulations. In fact, this bubble of high porosity is not fixed in time and space: the figure only shows a snapshot at one particular time. The bubble actually tends to move up until it reaches the free surface, then another bubble appears at the bottom in a cyclic manner - very much like air bubbles produced in a water tank. This is in clear contrast with the cavity regime in plot $(b)$ ($q^*=1.16\times10^{-3}$) in which the porosity field is stationary.

After setting the injection rate back to zero and reaching a final equilibrium state, a region of high porosity remains ($n\approx  0.42$ locally) above the injection point and throughout the layer (plot $(d)$ in figure \ref{new_figures:porosity_cavity}). It denotes an irreversible change of porosity in the layer after the injection steps.

\begin{figure}[H]
    \centering
        \includegraphics[width=10cm]{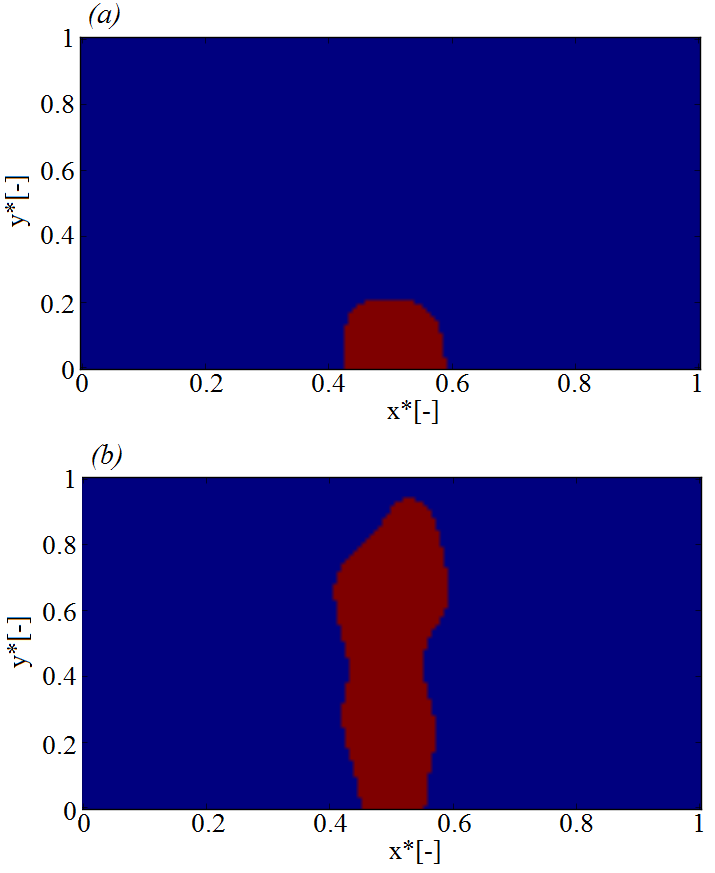}   
    \caption{Evolution of the fluidized zone for narrow aperture ($a^*$ = 0.1). Blue zone represents non-fluidized zone ($\dfrac{\sigma}{\sigma_{o}} \geq 0.1$). Red zone corresponds to fluidized zone  ($\dfrac{\sigma}{\sigma_{o}} < 0.1$).  (a) Cavity regime, $q^*$ = 0.00116. (b) Chimney regime, $q^*$ = 0.00145.}
    \label{new_figures:effective_stress_map_narrow}
\end{figure}

\begin{figure}[H]
    \centering
        \includegraphics[width=10cm]{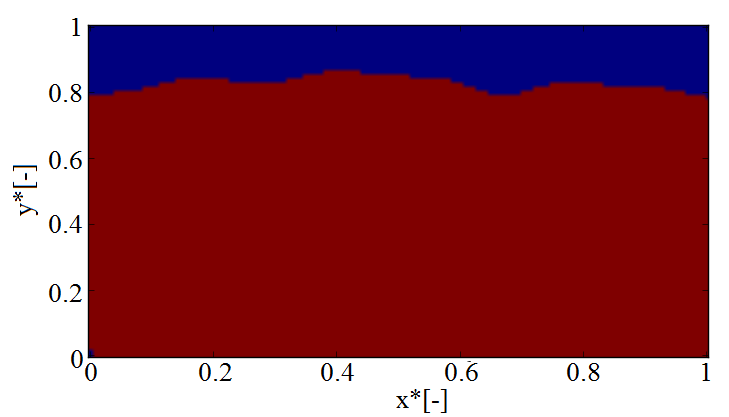}   
    \caption{Fluidized zone for a wide aperture (a* = 0.9) and $q^*$ = 0.00156, an important part of the sample has liquefied. Blue zone represents non-fluidized zone ($\dfrac{\sigma}{\sigma_{o}} \geq 0.1$). Red zone corresponds to fluidized zone  ($\dfrac{\sigma}{\sigma_{o}} < 0.1$).}
    \label{new_figures:effective_stress_map_wide}
\end{figure}

Cavities and chimneys may also be analyzed by means of the effective stress (Eq. \ref{eq:effectiveS}). However, null effective stress was never clearly found in the simulations since particle collisions occur in the fluidized zone and the corresponding contacts are reflected in the effective stress (arguably, zero-contact states hardly exist in agitated granular suspensions - see e.g. \cite{marzougui2015microscopic}). Conventionally, the fluidized zone in the simulations is defined by the points where the effective stress is less than 10$\%$ of the initial effective stress. This threshold is such that the cavity shape in plot $(b)$ in figure \ref{new_figures:porosity_cavity} matches the cavity in the top map in figure \ref{new_figures:effective_stress_map_narrow}. After some iterations we have found 10$\%$ of the initial effective stress is a good criterion to predict fluidized zones located inside the specimen. More restrictive thresholds (i.e. 5$\%$ of the initial effective stress) would delay the actual formation of the cavity and introduce some noise as effective stresses are never zero due to internal collisions. A coarse-grained function has been used as well when plotting the dimensionless effective stress maps in order to obtain accurate and consistent data. Therefore, the averaged effective stress zones leading to fluidization never reach, in appearance, the top of the layer. Furthermore, the height of the cavity can be found by means of the effective stress criterion as the highest fluidized point inside the sample. Cavity (plot $(a)$ in figure \ref{new_figures:effective_stress_map_narrow}) and chimney (plot $(b)$ in figure \ref{new_figures:effective_stress_map_narrow}) of fluidization can be easily identified when the injection area is small ($a^* = 0.1$). On the contrary, concerning large apertures (figure \ref{new_figures:effective_stress_map_wide}), fluidized zone does not look as a chimney due to the fact that an important part of the sample has liquefied.

\subsection{Comparison with experiments}\label{comparison2}

The results from the previous sections let the height of the cavity be defined by considering in both models the maximum height where the fluidization criterion is met.
It is thus possible to compare the models with the data from \cite{philippe2013localized}, given in terms of height of cavity versus flow rate.

Practically the fluidized zone in the tests was identified using time-averaged images of the granular layer, from which the regions with moving grains could be distinguished from the static ones according to a particular threshold. It was not possible to apply the very same definition of fluidization in the models, mainly because the details of the experimental steps and the thresholds were not reported. This possible cause of discrepancy has to be kept in mind.

In the models, the injection area has been defined so far as a band covering the whole depth of the sample while the laboratory tests were done with a smaller and circular orifice. For this comparison, the injection area has been modified in the DEM-PFV model to match that of the experiment. It is obviously not possible to reflect this particular geometry in the theoretical model, which is strictly two-dimensional.
Besides, the physical properties of the materials used in the test (table \ref{tab:sampleCharacteritics}) have been used directly as input parameters of both the theoretical and the numerical model. The only exception is the particle size: in the numerical model the grains are larger  than the real ones. We do not expect significant bias from this upscaling, as demonstrated in section \ref{Sensitivity_analysis}.

\begin{figure}[H]
    \centering
        \includegraphics[width=10cm]{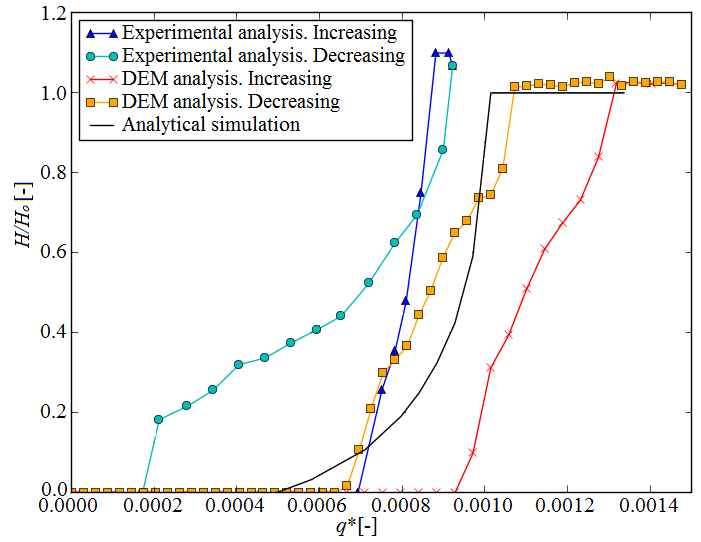}   
    \caption{Normalized height of the cavity in the theoretical and numerical models, compared with the experimental data from \cite{philippe2013localized}.}
    \label{new_figures:experimental_DEM_curves}
\end{figure}

Figure \ref{new_figures:experimental_DEM_curves} shows the ratio of cavity height $H_{c}$ to initial height of the specimen $H_{o}$. A qualitative agreement is found between the tests and the DEM-PFV simulations. The experimental increasing path is slightly steeper than the numerical one. Regarding the decreasing path, both experimental and numerical curves have the same tendency when $H_{c}/H_{o}$ >0.5. On the contrary, the gap between the decreasing curves widens when $H_{c}/H_{o}$ < 0.5, as the numerical curves is much steeper than the experimental one. We can clearly observe similar slopes when $H_{c}/H_{o}$ >0.5 for all the curves except for the experimental increasing path. This is related to the unknown initial packing from the experiments, which is presumably looser than the numerical samples and ease the propagation of the cavity. Besides, the numerical model overestimates the flux for cavity and chimney development (a shift is evidenced between numerical and experimental curves). This difference is mainly attributed to the fact that the permeability resulting from the numerical model is not exactly the same as the one of the real packing of glass beads. However, it is worth noting that permeability is not an input of the model but results from the description of solid fluid interactions at particle scales. Such interactions are described in the numerical model without any fitting parameter. In these conditions, numerical predictions may be considered satisfying from a quantitative point of view (it would be easy to fit very closely the experimental data by adding a single fitting parameter to the computation of the permeability in the model). The gap between the curves in the decreasing path when $H_{c}/H_{o}$ < 0.5 could be attributed to diameter differences. Experimental specimen height is $H \approx 24D$, where $D$ is the mean diameter. On the other hand, numerical sample height is $H \approx 10D$. Furthermore, experimental and DEM results may differ as a consequence of the different criteria used to obtain the height of the fluidized zone. 

The analytical solution has been included in figure \ref{new_figures:experimental_DEM_curves} as well. We can observe a smooth transition between the origin of the cavity and the point it reaches the top of the sample forming a chimney. Despite the fact analytical curve is near the experimental and numerical ones, analytical solution is based on a 2D media. On the contrary, numerical and laboratory simulations have a three-dimensional effect within the specimen due to the upward flow was injected through a circular orifice at the bottom of the bed rather than a rectangle covering the depth of the sample (see figure \ref{new_figures:studycase}).

\subsection{Flux-pressure relationship and fluidization regimes}\label{comparison}

\begin{figure}[ht]
    \centering
        \includegraphics[width=10cm]{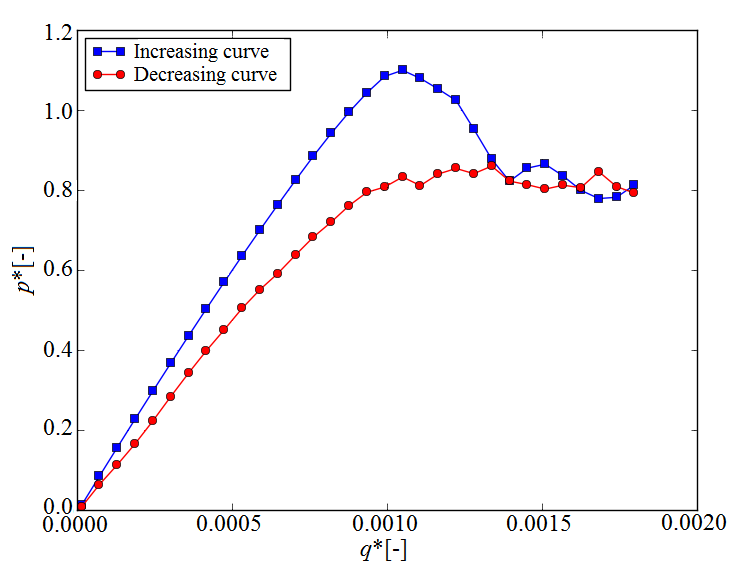}   
    \caption{Dimensionless pressure-flux, $p^*$-$q^*$, curve for an increasing-decreasing cycle of flux at the bottom of the sample. DEM-PFV simulation with an injection aperture $a^*$ = 0.1}
    \label{new_figures:pressure_inc_dec}
\end{figure}

In figure \ref{new_figures:pressure_inc_dec} the fluid pressure at the inlet in the DEM-PFV simulation is plotted as a function of the imposed flow rate. The imposed rate is increased over time up to a maximum, then decreased back to zero, with increments $\Delta q^*=0.000057$. Each flux value is kept constant over enough simulated time to exhibit the stationary solutions.

Very low flux values correspond to the situation when the particles do not move significantly and the pore pressure increases linearly with the flow rate, as expected from Eq. \ref{eq:Ec3.2}. At larger discharges the expansion regime leads to a noticeable increase of the porosity and hydraulic conductivity, such that the pressure is no longer proportional to the flux. Pore pressure keeps increasing until it reaches a peak ($p^* \approx 1.10$). As the peak value is reached, the fluidized cavity starts developing. Further increase of the flux results in a decreasing pressure as the cavity progressively approach the free surface. When the chimney is fully developed the pressure tends to a residual value in average, although the bubbling trend commented in previous section produces some fluctuation around that value (see figure \ref{new_figures:pressure_inc_dec}).

In the flux-decreasing phase, pressure values are always below the first flux-increasing curve. It is easily explained by the increased porosity leading to greater conductivity. The irreversible increment of porosity in the granular layer is evidenced in figure \ref{new_figures:porosity_cavity}$(d)$ . Starting from an initially dense material ($n \approx$ 0.365, close to the random close packing RCP), fluidization results in making looser the granular bed (at least locally, reaching porosity $n \approx$ 0.41 not so far from the random loose packing RLP). Then if the flux is increased again (not shown in the present work), no peak is observed in the $p^*$-$q^*$ plot, following closely the flux decreasing curve.

Hereafter, we focus on the increasing phase since one the main focus of this work is on the initiation and development of the fluidized zone.

\begin{figure}[ht]
    \centering
        \includegraphics[width=10cm]{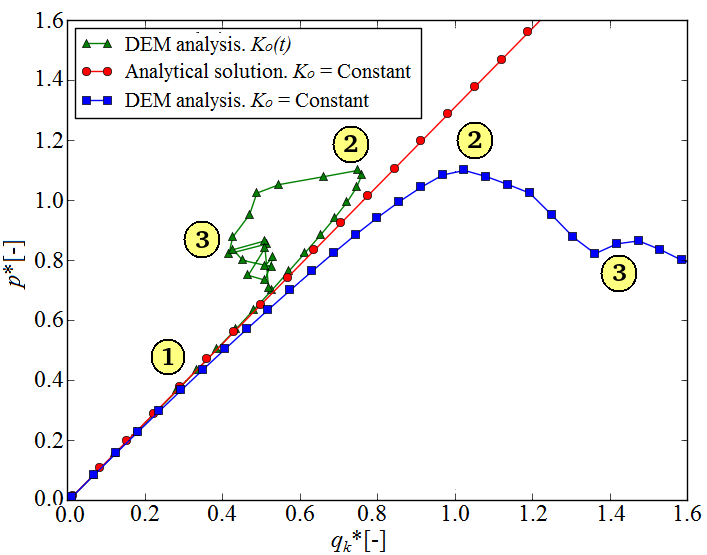}   
    \caption{$p^*$-$q^*_k$ curves comparison at the center of the injection orifice with an injection aperture $a^*$ = 0.1.}
    \label{new_figures:pressure_comparison}
\end{figure}

The non-linearity of the flux-pressure relation before the peak can be explained by the increase of the hydraulic conductivity as porosity increases in the expansion regime. In order to isolate this effect the results can be plotted considering the second definition of dimensionless flux $q^*_{k}$ (see Eq.~\ref{eq:Ec5}), where the updated conductivity is used. The hydraulic conductivity at one particular time is obtained based on Eq.\ref{eq:Ec5} and considering the average porosity above the injection area. In figure \ref{new_figures:pressure_comparison} three curves are considered: the analytical solution (red curve with circle symbols), the numerical results interpreted as if conductivity was constant in time (blue line with square symbols) and the interpretation including the updated conductivity (green line with triangle symbols). 


Figure \ref{new_figures:pressure_comparison} shows a perfect fit of the analytical solution by the numerical results at low fluxes ("0"-"1" path) when plotted with the updated conductivity. After point "1" the granular assembly starts expanding significantly and some non-linearity appears but it is less significant in the $p^*-q^*_{k}$ plot. 

The cavity regime begins immediately after the pressure peak is reached (point "2"). As can be seen in figure \ref{new_figures:pressure_comparison}, the blue-squared $p^*-q^*_{k}$ curve clearly deviates from the analytical solution after this point, reaching lower values of $q^*_k$. This is to be understood as a deviation from the homogeneous porosity field assumed for the theoretical derivation: the conductivity increases much more above and around the inlet than in the rest of the simulated layer. Finally, the chimney regime is reached (point "3" in figure \ref{new_figures:pressure_comparison}), triggering bubbling events through the layer. The non-steady nature of the chimney is visible in the scattering of the results after point "3" in terms of both $p^*$ and $q_k^*$. 

\begin{figure}[ht]
    \centering
        \includegraphics[width=10cm]{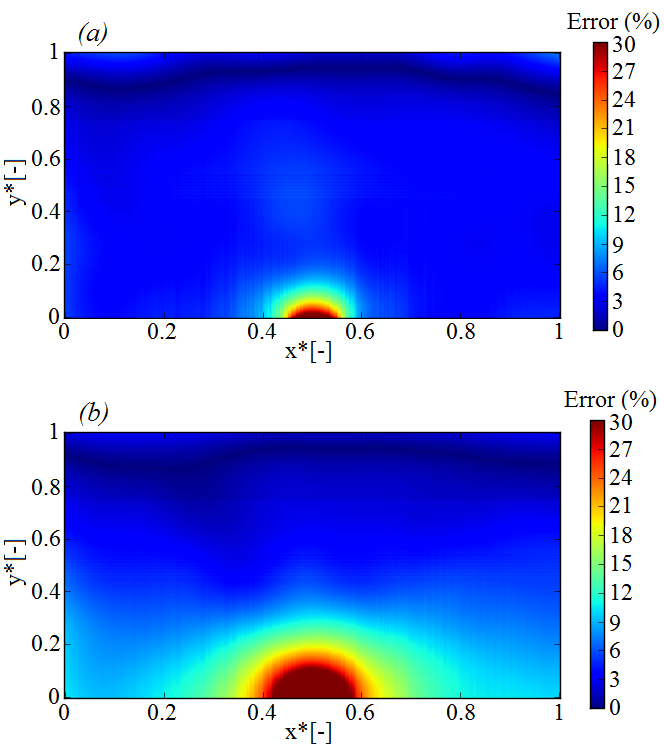}   
    \caption{Relative error between numerical and analytical pressure within the entire sample. $(a)$ Expansion regime, $q^*$ = 0.00095.  $(a)$ Cavity regime, $q^*$ = 0.00116.}
    \label{new_figures:pressure_map}
\end{figure}

In order to compare the numerical and the analytical results in more details the difference between the pressure fields is shown in figure \ref{new_figures:pressure_map}. The difference is normalized by the analytical pressure at the inlet.

Figure \ref{new_figures:pressure_map} evidences that the error between the pressure obtained numerically and the analytical solution is especially high in the vicinity of the inlet, where error reaches values up to 30$\%$. However, error rapidly decreases far from the injection point where the average error is usually lower than 10$\%$ as permeabilities do not significantly change. Moreover, error increases within the entire specimen when fluidization begins, as we can see in plot $(b)$ of figure \ref{new_figures:pressure_map}, though the error is still low and acceptable.

\begin{figure*}[ht]
    \centering
        \includegraphics[width=17cm]{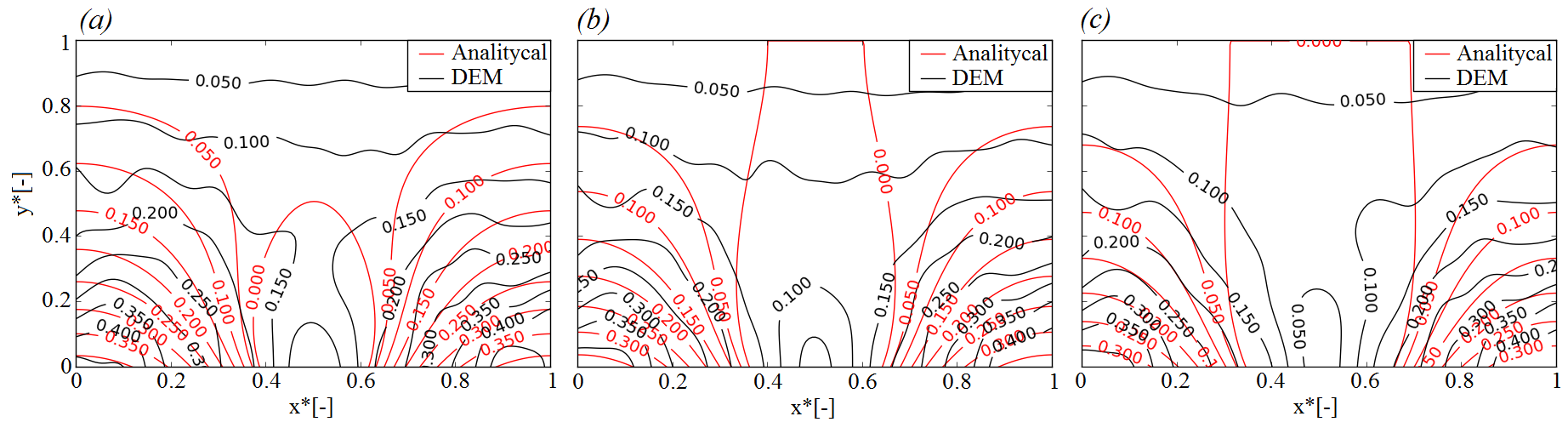}   
    \caption{Evolution of the analytical and numerical effective stress fields for different flow rates (injection aperture $a^*$ = 0.1). $(a)$ $q^*$ = 0.00105: cavity regime; $(b)$ $q^*$ = 0.00114: chimney regime in the theoretical model; cavity regime in the numerical model. $(c)$ $q^*$ = 0.00120: chimney regime.}
    \label{new_figures:effective_comparison}
\end{figure*}

The shape of the fluidized zone resulting from the analytical and the numerical simulations can be compared through the map of effective stress. The analytical fluidized zone is defined by the null-pressure isoline, in the simulation we retain the $\sigma'^{*} = 0.1$ isoline as discussed previously. Figure \ref{new_figures:effective_comparison} shows that the isolines are relatively similar in the undisturbed regions on both sides of the sample. On the contrary, they differ significantly above the inlet. As a matter of fact, a chimney-shaped fluidized zone appears in the analytical solution in the middle image ($q^* = 1.14\times10^{-3}$) while at the same flow rate the cavity only starts developing in the simulation. The chimney regime is attained for a value of $q^* = 1.20 \times10^{-3}$ in the numerical simulation, in this situation, $\sigma'^{*} < 0.1$ is found in every point located vertically between the free surface and the inlet.

\section{Sensitivity analysis}\label{Sensitivity_analysis}

\begin{figure*}[ht]
    \centering
        \includegraphics[width=16cm]{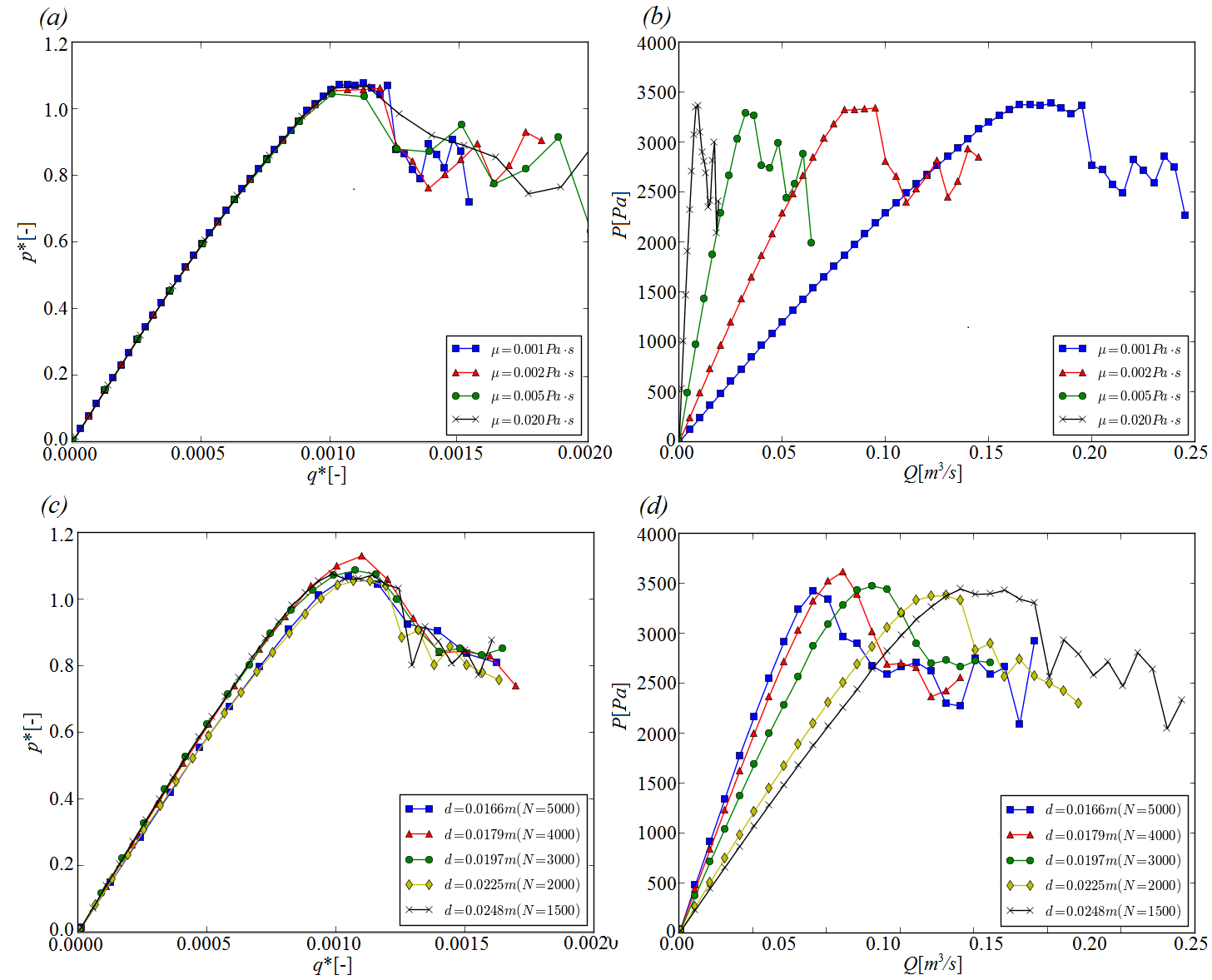}   
    \caption{Sensitivity analysis on pressure-flux curves with injection aperture $a^*$ = 0.1. Viscosity dependency with normalized $p^*$-$q^*$ variables $(a)$ and physical units $(b)$; diameter dependency with normalized $p^*$-$q^*$ variables $(c)$ and physical units $(d)$.}
    \label{new_figures:sensitivity_curves}
\end{figure*}

Provided that the normalized pressure ($p^*$) and flux ($q^*$) given in Eq. \ref{eq:Ec6} and \ref{eq:Ec4} are relevant dimensionless variables, there should be a unique relationship between them independently of the mean particle size and fluid viscosity. This is verified in this section. Figure \ref{new_figures:sensitivity_curves} shows the pressure-flux relations for a monotonously increasing flux and $a^*$ = 0.1, in terms of both the physical units ($P-Q$) and the dimensionless quantities ($p^*$-$q^*$). For fluid viscosity ranging from $\mu = 10^{-3}$ to $2 \cdot 10^{-2} Pa \cdot s$, and mean particle size ranging from $d$ = 1.66 to 2.48 $cm$. The pressure-flux results are collapsed in one single curve when expressed with $p^*$ and $q^*$, which confirms the relevance of the chosen variables and validates the proposed dimensionless numbers. Besides, the dimensional plots (Figure \ref{new_figures:sensitivity_curves}, diagrams on the right) all show a marked peak at approximately $P=3400Pa$ ($p^*=1.05$) regardless of the flow rate. This value corresponds to pressure gradients balancing gravitational and frictional forces. This suggests that fluid pressure, instead of flow rate, is the most natural parameter for defining fluidization criteria.

\begin{figure}[H]
    \centering
        \includegraphics[width=10cm]{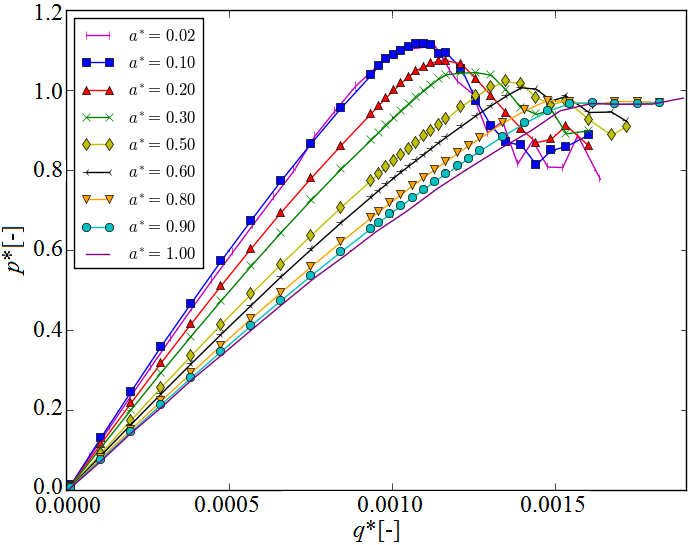}   
    \caption{Aperture dependency. $p^*$-$q^*$ curves at the bottom of the sample with different injection apertures.}
    \label{new_figures:dimensionless_aperture_curves}
\end{figure}

Nevertheless, the $p^*$-$q^*$ relation is not absolutely unique as it depends on the remaining dimensionless variable: normalized aperture of the injection area. This dependency is shown in Figure \ref{new_figures:dimensionless_aperture_curves} for $a^*$ ranging from 0.02 to 1 ($a^*=1$ means uniform influx through the surface of an infinite half-space).

A clear pattern can be identified in the figure \ref{new_figures:dimensionless_aperture_curves}. Large apertures induce liquefaction over all the granular medium when the pressure peak attains a value close to $p^* = 1$, the expected value for homogeneous fluidization.

On the other hand, the excess of pore pressure required to reach fluidization is larger for small apertures ($p^* \approx 1.1$). The reason why $p^*$ exceeds unity with small apertures is that the mobility of the column above the injection zone is constrained not only by gravity (gravity alone would lead to $p^*=1$ as an upper bound) but also by interactions with particles on both sides of the column. The pore pressure required to develop a cavity and a chimney of fluidization in the specimen must then exceed the sum of weight and a downward force coming from contact interactions between the stable mass and the mobile particles. As aperture becomes larger the mass of stable particles is progressively reduced and it eventually disappears in homogeneous fluidization ($a^*\rightarrow 1$), hence the additional downward force vanishes and no peak is observed. 

\begin{figure}[H]
    \centering
        \includegraphics[width=9cm]{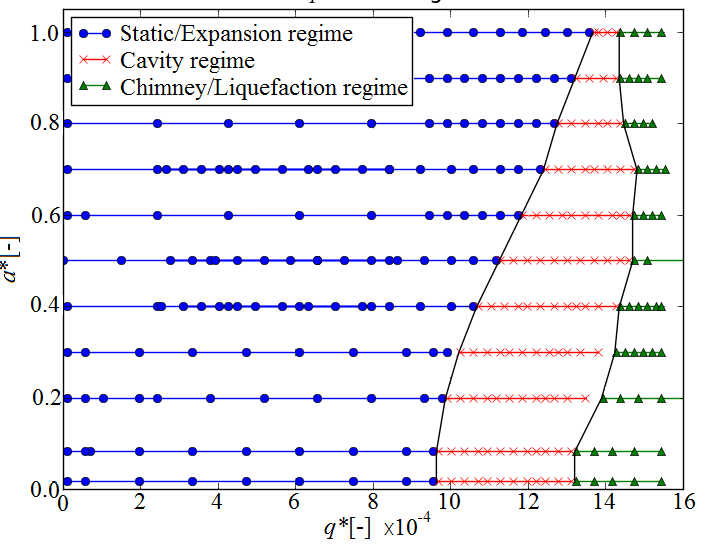}   
    \caption{Occurence of the different fluidization regimes (static/expansion, cavity or chimney) depending on aperture $a^*$.}
    \label{new_figures:DEM_diagram}
\end{figure}

The aperture dependency is summarized in the diagram of figure \ref{new_figures:DEM_diagram} where the different fluidization regimes are identified for particular combinations of simulated flow rate and aperture. As mentioned before and confirmed in figure \ref{new_figures:DEM_diagram}, large apertures ($a^*$ close to 1) lead to the entire liquefaction of the granular assembly rather than forming a cavity and a chimney of fluidization (cavity regime gets narrower near $a^*=1$.).

\begin{figure}[H]
    \centering
        \includegraphics[width=9cm]{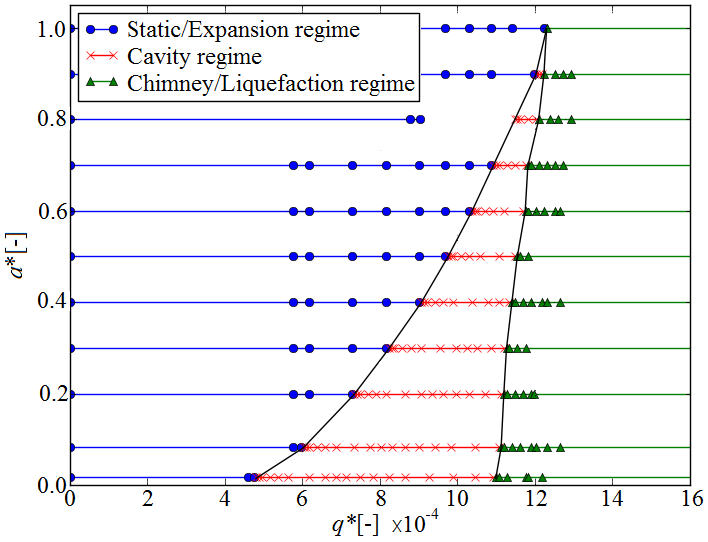}   
    \caption{Theoretical prediction of the different regimes (static/expansion, cavity or chimney regimes) in the dimensionless aperture-flux, $a^*$-$q^*$, plane.}
    \label{new_figures:analytical_diagram}
\end{figure}

Qualitatively, similar tendencies are predicted by Eq. \ref{eq:Ec3.2} of the theoretical model (see also Figure \ref{new_figures:effective_field}). The regimes deduced in this way are presented in figure \ref{new_figures:analytical_diagram}. Nevertheless, three significant differences are found. Firstly, the transitions between the different regimes are shifted to lower fluxes compared to the simulations since the increase of hydraulic conductivity induced by the injection is not accounted for in Eq. \ref{eq:Ec3.2}, which thus overestimates the actual pore pressure for a given flux and anticipate the fluidization.

Secondly, the line separating the expansion and the cavity regimes for narrow apertures is nearly vertical in the former diagram while it is getting horizontal when the aperture decreases. This is due to the divergence of the theoretical inlet pressure when $a^*\rightarrow 0$, which results in a finite sized cavity for every non-zero influx in this limit. It does not occur in the numerical simulations because the spatial discretization of the flow problem introduces a lower bound for the range of effective aperture: when $a^*$ is of the order of the average distance between two solid particles further decrease in $a^*$ means no change in the solution. Likewise, we would not expect any difference if the injection was done with syringes using needles of different sizes, as soon as the needles are smaller than the pores of the material.

Finally, the transition between the cavity and the chimney regimes has a "D" shape in the simulation (maximum $q^*$ values for $a^* = 0.5$ - $0.6$, and $q^*$ values decrease when narrower or wider apertures are considered) whereas it is approximately a straight line theoretically. The shift toward lower $q^*$ values for wide apertures ($a^*$ near 1) is expected because no interaction remains between fluidized and non-fluidized zones, then fluidization is reached for lower flux. As a matter of fact, this phenomenon can be observed in figure \ref{new_figures:dimensionless_aperture_curves}. In $a^*=0.80$, $a^*=0.90 $ and $a^*=1.00 $ cases, liquefaction starts when the plateau is reached for $q^* \approx 1.5 \times10^{-3}$. In $a^*=0.50$ and $a^*=0.60$ cases, chimney of fluidization is attained for slightly larger $q^* \approx 1.65 \times10^{-3}$ (after this point pressure values oscillate as a consequence of the bubbling effect).

\section{Conclusions}
In this work, fluidization of a granular bed through an injection orifice has been investigated numerically using DEM-PFV simulations and analytically using a simplified continuum model.

Qualitatively the DEM simulations were found to compare well with available data in terms of the different fluidization regimes. To some extent a reasonable quantitative agreement was also found, without introducing any fitting parameter, although some differences remain. The origin of these differences is unclear as long as no published data include pore pressure measurements. 

The theoretical model was assuming a constant hydraulic conductivity in space and time and the effective stress was used for defining a fluidization criterion. It could approximate the DEM solution quite well at low fluxes. It deviates from the DEM solution at larger fluxes when expansion of the bed and localized fluidization leads to significant changes of the conductivity in space and time. Nevertheless, the analytical model predicts the transitions between the different regimes and it provides a simple framework to explain the main trends. This theoretical approach may also shed light on the interactions between two or more adjacent chimneys in the case of multiple injection points \cite{ngomainteraction,philippe2013localized}. 

The numerical model allowed to describe the internal configuration by means of the effective stress and the porosity fields within the granular medium. This approach enabled to define accurately the cavity and chimney shape. Besides, compression-decompression cycles were evidenced once the chimney regime was reached.

The size of the injection area was found to determine whether or not a chimney regime exists for certain injection rates. Small injection areas lead to an early regime of fluidization with respect to the injected flux. In this case, gradual increase of the injection rate results in a peak of the inlet pressure at the transition between the cavity regime and the chimney regime. On the contrary, large apertures produce liquefaction for larger injection rates, the chimney is wider and difficult to identify due to liquefaction occurring over most of the sample instead of a localized zone. Consistently, the peak in the pressure-flux curves tend to disappear and the evolution of pressure increases monotonically as a function of the injection rate until it reaches a plateau. The domains of occurrence of the different regimes have been defined in terms of dimensionless aperture and dimensionless flux.

Our analysis suggests that the inlet pressure is the primary variable controlling fluidization more directly than the injection rate. The published data reports the injection rate only, hence the difficulty in comparing experiments and simulations.

Our suggestion for future experiments is to measure the inlet pressure, ideally. Alternatively, an independent measurement of the permeability would enable the calculation of inlet pressure at least for the initial stages before any significant change of the local porosity. 

\section{Acknowledgement}
This work has been partially supported by the LabEx Tec 21 (Investissements d’Avenir - grant agreement $n^o$ ANR-11-LABX-0030) 

\newpage
\newpage
\appendix
\section{\\Appendix: Analytical expression of the pressure field induced by a single injection point} \label{App:Appendix}

In order to obtain analytical expressions of the pore pressure the following assumptions have been considered, leading to a standard Laplace problem:
\begin{itemize}

  \item  The fluid flows through a macroscopically homogeneous porous medium.
  \item  The flow velocity follows Darcy's law, its velocity $v$ is proportional to the local pressure gradient: \begin{equation}\label{eq:Ec9}  v=\dfrac{k}{\mu }\nabla P \phantom{\hspace{3cm}} \end{equation} where $k$ is the intrinsic permeability and $\mu$ the dynamic viscosity (both uniform in space).
  \item  The fluid is incompressible, hence a divergence-free condition: $\nabla \cdot v = \dfrac{k}{\mu}\nabla^{2}P=0$ everywhere except at injection points.
\end{itemize}

\begin{figure}[ht]
    \centering
        \includegraphics[width=7cm]{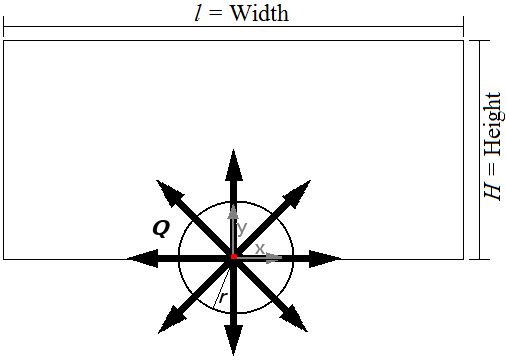}   
    \caption{Fluid flowing outwards in a 2D porous media due to a punctual source.}
    \label{new_figures:analytical_idea}
\end{figure}

We first define the solution corresponding to a punctual source in an infinite domain. The flow has a rate $Q[m^{3}/s]$ (see figure \ref{new_figures:analytical_idea}) and the radial component of the velocity of the fluid at a distance $r$ can be expressed as: \begin{equation} \label{eq:Ec10}  v=\dfrac{Q}{2 \pi r} \phantom{\hspace{3cm}} \end{equation}

Combining and solving Eq. (\ref{eq:Ec9}) and (\ref{eq:Ec10}) for a punctual source leads to the expression of the pressure drop $\Delta P$ between two points located at distances $r$ and $r_{o}$ respectively from the punctual source:
\begin{equation}\label{eq:Ec11}  \Delta P = P_{o} - P = \dfrac{\mu Q}{2 \pi k}\cdot ln \left( \dfrac{r}{r_{o}} \right) = \dfrac{\mu Q}{2 \pi k}\cdot ln \left( \dfrac{\sqrt{x^{2}+y^{2}}}{r_{o}} \right)\end{equation}

This solution satisfies Laplace equation $( \bigtriangledown^{2} P=0)$ in an infinite 2D space. Due to the linearity of Laplace equation two or more potential functions of this form centered in different points can be combined to describe the pressure/flow fields induced by a set of punctual sources (where $Q$ can be negative or positive). Namely, an appropriate set of sources in an infinite medium can replicate the features of the pressure field induced by one single source in a finite sized domain. A no-flux condition at a boundary of the finite domain corresponds to a symmetry of the sources in the infinite domain. On the other hand, a null pressure condition corresponds to a skew-symmetry of the sources with respect to this boundary.

For our particular boundary conditions, the null pressure condition on the top boundary ($y=c/2$) suggests two conjugate sources with fluxes $Q$ and $-Q$ located on either side of this boundary at heights $y=0$ and $y=c$ (figure \ref{new_figures:infinite_net2}). However, the symmetry with respect to the no-flux bottom boundary ($y=0$) imposes to replicate the $-Q$ source in $y=-c$. In turns, this this replicate is itself reflected by a skew symmetric $Q$ source in $y=2c$, etc. Recursively, it leads to an infinite series of sources aligned vertically and alternating positive and negative fluxes. In addition, the no flux conditions on the lateral boundaries leads to symmetries with respect to the lines $x=-l/2$ and $x=l/2$. The previous set of conjugate sources is thus replicated periodically in the horizontal direction. Finally, the pressure at any point of the finite domain can be obtain by assuming the pressures associated to each source of the infinite set of sources/sinks:

\begin{figure}[ht]
    \centering
        \includegraphics[width=10cm]{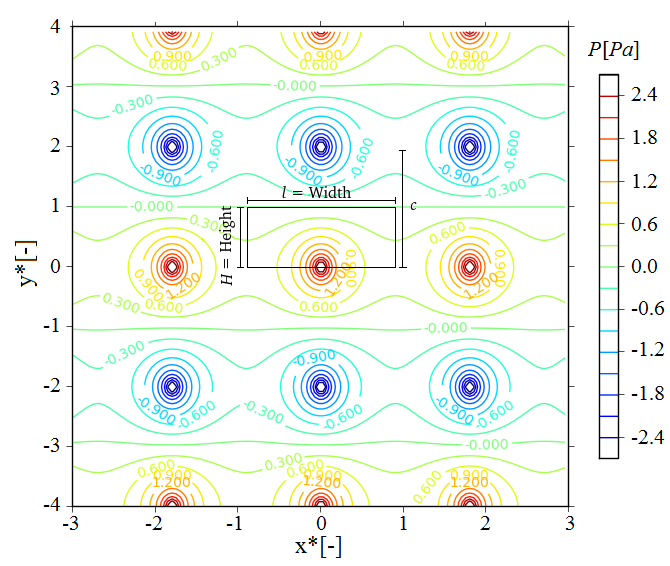}   
    \caption{Pressure field for semi-infinite sources net.}
    \label{new_figures:infinite_net2}
\end{figure}

 \thickmuskip=0mu
 \makeatletter
 \def\@eqnnum{{\normalsize \normalcolor (\theequation)}}
  \makeatother
 { \small \begin{equation}\label{eq:Ec3}  P=\dfrac{Q\mu}{2 \pi k} \sum\limits_{j=-\infty}^\infty \sum\limits_{i=-\infty}^\infty -1^{|j|} \left[ln(\sqrt{(x-i\ l)^{2}+(y-jc)^{2}}) \right]
 \end{equation} }
where $c = 2 \cdot H$, as sources and sinks are separated by a distance which is twice the height of the thickness of the layer. Figure \ref{new_figures:infinite_net2} shows the pressure field induced by such an array of sources, the size of the actual finite domain is superimposed.

\bibliographystyle{unsrt}
\bibliography{biblucsi}

\begin{thebibliography}{10}

\bibitem{payne2008remediation}
Fred~C Payne, Joseph~A Quinnan, and Scott~T Potter.
\newblock {\em Remediation hydraulics}.
\newblock CRC Press, 2008.

\bibitem{peng1997hydrodynamic}
Yimin Peng and LT~Fan.
\newblock Hydrodynamic characteristics of fluidization in liquid-solid tapered
  beds.
\newblock {\em Chemical Engineering Science}, 52(14):2277--2290, 1997.

\bibitem{weisman1994design}
Richard~N Weisman and Gerard~P Lennon.
\newblock Design of fluidizer systems for coastal environment.
\newblock {\em Journal of waterway, port, coastal, and ocean engineering},
  120(5):468--487, 1994.

\bibitem{bonelli2013erosion}
St{\'e}phane Bonelli.
\newblock {\em Erosion in geomechanics applied to dams and levees}.
\newblock John Wiley \& Sons, 2013.

\bibitem{foster2000statistics}
Mark Foster, Robin Fell, and Matt Spannagle.
\newblock The statistics of embankment dam failures and accidents.
\newblock {\em Canadian Geotechnical Journal}, 37(5):1000--1024, 2000.

\bibitem{fry1997erosion}
JJ~Fry, G~Degoutte, and A~Goubet.
\newblock L’{\'e}rosion interne: typologie, d{\'e}tection et r{\'e}paration.
\newblock {\em Barrages \& r{\'e}servoirs}, 6:126, 1997.

\bibitem{schaafsma2006investigation}
SH~Schaafsma, T~Marx, and AC~Hoffmann.
\newblock Investigation of the particle flowpattern and segregation in tapered
  fluidized bed granulators.
\newblock {\em Chemical Engineering Science}, 61(14):4467--4475, 2006.

\bibitem{sutar2012mixing}
Harekrushna Sutar and Chandan~Kumar Das.
\newblock Mixing and segregation characteristics of binary granular material in
  tapered fluidized bed: A cfd study.
\newblock 2012.

\bibitem{sutkar2013spout}
Vinayak~S Sutkar, Niels~G Deen, and JAM Kuipers.
\newblock Spout fluidized beds: Recent advances in experimental and numerical
  studies.
\newblock {\em Chemical Engineering Science}, 86:124--136, 2013.

\bibitem{soderlund2007evaluating}
M.~S{\"o}derlund, P.~Bots, P.~Eriksson, P.~Nilsson, and J.~Hartlen.
\newblock Evaluating the domino-effect of failure of critical constructions due
  to damage of underground water pipelines.
\newblock {\em Loss Prevention Bulletin--Institution of Chemical Engineers},
  195:22--27, 2007.

\bibitem{briens1997characterization}
LA~Briens, CL~Briens, A~Margaritis, SL~Cooke, and MA~Bergougnou.
\newblock Characterization of channelling in multiphase systems. application to
  a liquid fluidized bed of angular biobone particles.
\newblock {\em Powder technology}, 91(1):1--9, 1997.

\bibitem{gallo2004steady}
F~Gallo and AW~Woods.
\newblock On steady homogeneous sand--water flows in a vertical conduit.
\newblock {\em Sedimentology}, 51(2):195--210, 2004.

\bibitem{kohl2014magnetic}
MH~K{\"o}hl, G~Lu, JR~Third, Klaas~P Pruessmann, and CR~M{\"u}ller.
\newblock Magnetic resonance imaging (mri) of jet height hysteresis in packed
  beds.
\newblock {\em Chemical Engineering Science}, 109:276--283, 2014.

\bibitem{philippe2013localized}
P.~Philippe and M.~Badiane.
\newblock Localized fluidization in a granular medium.
\newblock {\em Physical Review E}, 87(4):042206, 2013.

\bibitem{zoueshtiagh2007effect}
Farzam Zoueshtiagh and Alain Merlen.
\newblock Effect of a vertically flowing water jet underneath a granular bed.
\newblock {\em Physical Review E}, 75(5):056313, 2007.

\bibitem{anderson1967fluid}
T~Bo Anderson and Roy Jackson.
\newblock Fluid mechanical description of fluidized beds. equations of motion.
\newblock {\em Industrial \& Engineering Chemistry Fundamentals},
  6(4):527--539, 1967.

\bibitem{gidaspow1991hydrodynamics}
Dimitri Gidaspow, Rukmini Bezburuah, and J~Ding.
\newblock Hydrodynamics of circulating fluidized beds: kinetic theory approach.
\newblock Technical report, Illinois Inst. of Tech., Chicago, IL (United
  States). Dept. of Chemical Engineering, 1991.

\bibitem{mickley1955mechanism}
HS~Mickley and Do~F Fairbanks.
\newblock Mechanism of heat transfer to fluidized beds.
\newblock {\em AIChE Journal}, 1(3):374--384, 1955.

\bibitem{cui2014coupled}
Xilin Cui, Jun Li, Andrew Chan, and David Chapman.
\newblock Coupled {DEM-LBM} simulation of internal fluidisation induced by a
  leaking pipe.
\newblock {\em Powder Technology}, 254:299--306, 2014.

\bibitem{ngomainteraction}
Jeff Ngoma, P~Philippe, S~Bonelli, JY~Delenne, and F~Radjai.
\newblock Interaction between two localized fluidization cavities in granular
  media: Experiments and numerical simulation.
\newblock {\em Geomechanics from Micro to Macro: Proc. of the Int. Symposium on
  Geomechanics from Micro to Macro – IS-Cambridge 2014}, pages 1571--1576,
  1-3 Sept 2014.
\newblock Cambridge, UK.

\bibitem{catalano2014pore}
Emanuele Catalano, Bruno Chareyre, and Eric Barth{\'e}lemy.
\newblock Pore-scale modeling of fluid-particles interaction and emerging
  poromechanical effects.
\newblock {\em International Journal for Numerical and Analytical Methods in
  Geomechanics}, 38(1):51--71, 2014.

\bibitem{chareyre2012pore}
Bruno Chareyre, Andrea Cortis, Emanuele Catalano, and Eric Barth{\'e}lemy.
\newblock Pore-scale modeling of viscous flow and induced forces in dense
  sphere packings.
\newblock {\em Transport in porous media}, 94(2):595--615, 2012.

\bibitem{chen1998lattice}
Shiyi Chen and Gary~D Doolen.
\newblock Lattice boltzmann method for fluid flows.
\newblock {\em Annual review of fluid mechanics}, 30(1):329--364, 1998.

\bibitem{cui20122d}
Xilin Cui, Jun Li, Andrew Chan, and David Chapman.
\newblock A 2d dem--lbm study on soil behaviour due to locally injected fluid.
\newblock {\em Particuology}, 10(2):242--252, 2012.

\bibitem{cui2013numerical}
Xilin Cui.
\newblock {\em Numerical simulation of internal fluidisation and cavity
  evolution due to a leaking pipe using the coupled DEM-LBM technique}.
\newblock PhD thesis, University of Birmingham, 2013.

\bibitem{catalano2011pore}
E.~Catalano, B.~Chareyre, A.~Cortis, and E.~Barth{\'e}l{\'e}my.
\newblock A pore-scale hydro-mechanical coupled model for geomaterials.
\newblock In {\em Particles 2011 II International Conference on Particle-Based
  Methods}, 2011.

\bibitem{yade:doc2}
{\v{S}}milauer et~al.
\newblock Dem formulation.
\newblock {\em Yade documentation 2nd ed. The Yade Project}, 2015.

\bibitem{marzougui2015microscopic}
Donia Marzougui, Bruno Chareyre, and Julien Chauchat.
\newblock Microscopic origins of shear stress in dense fluid--grain mixtures.
\newblock {\em Granular Matter}, 17(3):297--309, 2015.

\end{thebibliography}

\end{document}